\documentclass[12pt]{iopart}

\pdfoutput=1

\usepackage{iopams}
\usepackage{setstack}
\usepackage{color}
\usepackage{graphicx}
\usepackage{cite}

\begin{document}

\title[Unidirectional laning and migrating clusters of confined self-propelled particles]{Unidirectional laning and migrating cluster crystals in confined self-propelled particle systems}

\author{A. M. Menzel}

\address{
  Institut f\"ur Theoretische Physik II: Weiche Materie, Heinrich-Heine-Universit\"at D\"usseldorf, 
    Universit\"atsstra{\ss}e 1, D-40225 D\"usseldorf, Germany
}
\ead{menzel@thphy.uni-duesseldorf.de}

\pacs{87.18.Gh, 64.60.Cn, 61.30.Hn}

\begin{abstract}
One standard approach to describe the collective behaviour of self-propelled particles is the Vicsek model: point-like self-propelled particles tend to align their migration directions to the ones of their nearer neighbours at each time-step. Here we use a variant of the Vicsek model that includes pairwise repulsive interactions. 
Confining the system between parallel walls can qualitatively change its appearance: a laning state can emerge that is different from the ones previously reported. 
All lanes show on average the {\em same} migration direction of the contained particles with a {\em finite} separation distance between the lanes. 
Furthermore, in certain parameter ranges we observe collectively migrating clusters that arrange in an approximately hexagonal way. We suggest that the mechanism behind these regular textures is an overreaction in the alignment mechanism. Considering the more realistic scenario of non-point-like particles in the presence of confining surfaces is generally important for the comparison to experimental systems.
\end{abstract}

\maketitle

\section{Introduction}

The study of self-propelled particles is to a big extent inspired by nature. Bacteria developed different propulsion mechanisms to swim through a fluid environment \cite{darnton2007torque} or to migrate on surfaces \cite{
harshey2003bacterial,kearns2010field}. Likewise amoeba and tissue cells have the ability to crawl on substrates \cite{rappel1999self,szabo2006phase}. Although they form the smallest living biological units, the physics of these single cells is already very complex. Therefore, to understand the underlying principles of collective motion when many self-propelled objects act together, it is helpful to also investigate the behaviour of simple artificially generated self-propelled particles. 

For this purpose Janus particles that intrinsically break the forward-backward symmetry due to their two opposing different surfaces can be used. If one of the two surfaces catalyses a chemical reaction in the surrounding medium \cite{theurkauff2012dynamic} or is selectively heated by external illumination \cite{volpe2011microswimmers,buttinoni2012active}, propulsion of the Janus particles can result. Through spontaneous symmetry breaking also initially symmetric droplets in a liquid environment can start to self-propel due to chemical reactions on their surfaces \cite{thutupalli2011swarming}. 

The motion of granular hoppers on a vibrating substrate \cite{narayan2007long,deseigne2012vibrated} can be studied in the plane of the system, as long as the particles do not jump over each other. 
Other objects propelling in two spatial dimensions are droplets on surfaces \cite{nagai2005mode}. After a spontaneous symmetry breaking, these droplets migrate due to gradients of surface tension. Recently, the motion of surface droplets that are driven by  internally propagating reaction-diffusion waves has been observed \cite{kitahata2011spontaneous,kitahata2012spontaneous}. 

One route of modeling the collective behaviour of such particles consists of assigning to each object an active drive and letting them pairwisely interact by a repulsive potential. In this way, a manifold of different active collective states can be observed: disordered, jammed, clustering, crystalline, turbulent, bio-nematic, swarming, laning \cite{henkes2011active, mccandlish2012spontaneous, wensink2012meso, wensink2012emergent, fily2012athermal, bialke2012crystallization, menzel2012soft}. The laning state in a system of identical self-propelled particles \cite{wensink2012meso, wensink2012emergent, menzel2012soft} is of particular interest when we consider the orientational order of the migration directions. Within each lane, the self-propulsion directions are ordered in a polar way. That is, all particles within one lane migrate into the same direction on average. However, the global order is only nematic. This is because particles in neighbouring lanes propel into opposite directions. Such a state can be observed at high density and large aspect ratio of the particles when we start from random initial conditions for the velocity orientations \cite{wensink2012meso, wensink2012emergent, menzel2012soft}. 

Using a much more simplified approach, the onset of collective motion in self-propelled particle crowds was studied by Vicsek et al.\ \cite{vicsek1995novel}. In this case there are no potential interactions between the consequently point-like particles. Instead, at each time-step, polar alignment rules are applied to the velocity vectors: each particle averages over all the velocity orientations within a spherical environment around itself and matches its velocity direction accordingly. Competing orientational noise disturbs this velocity alignment. When the density is increased, or the noise intensity is decreased, a transition occurs from disordered motion to globally ordered polar collective motion. In the latter case all particles propel on average into the same direction. The velocity magnitude is assumed identical for each particle. 

Since the particles are considered as point-like in the Vicsek model, the local density can become very large. Spatial spreading results mainly from orientational noise, not from particle repulsion. Here we resolve this issue by introducing an additional velocity alignment rule that effectively leads to spatial avoidance. In this way the local density becomes bounded, but we keep the spirit of the Vicsek model of interactions based solely on alignment rules. We consider such a system in two spatial dimensions confined in a channel between two surfaces that repel particles again by velocity alignment rules. In certain parameter regimes we find a surface-supported laning scenario that is different from the previously reported cases \cite{mccandlish2012spontaneous, wensink2012meso, wensink2012emergent, menzel2012soft} and to our knowledge has not been observed before: first, all particles of the system on average migrate into the {\em same} direction; second, there is a {\em finite} gap between the lanes, its thickness being of the size of the lane thickness or larger. 
We therefore introduced the term unidirectional laning to characterise this state. For other parameter values we find the formation of migrating clusters. Since these clusters arrange in an approximately hexagonal order and collectively migrate into the same direction, we call this state a migrating cluster crystal. We consider our observations important concerning comparisons between the Vicsek model and real experiments because the particle density in real systems is bounded and an experimental system is usually confined by surfaces. 

First we will introduce the model (section \ref{alignment}) and then illustrate the impact of the pairwisely repulsive interaction (section \ref{repulsive}). An example for the phase behaviour is presented afterwards as a function of the channel width and the total particle density (section \ref{phases}). We then illustrate the new state of unidirectional laning (section \ref{laning}) and the mechanism behind it (section \ref{mechanism}). 
Some remarks on the role of the strength of the stochastic force are included (section \ref{noiserole}), 
as well as the observation of the migrating cluster crystals (section \ref{clusters}). The role of the pairwisely repulsive interaction is briefly discussed (section \ref{role}) before we conclude (section \ref{conclusion}).

\section{Alignment rules}\label{alignment}

In the following we consider $N$ particles labeled by an index $i$. All of them self-propel with a constant migration speed $u$ in the two-dimensional plane. The vector $\mathbf{r}_i=\left(x_i,y_i\right)$ gives the position of the $i$th particle and the angle $\theta_i$ parameterises the orientation of its migration direction. At discrete times $t$ the increments $\Delta \mathbf{r}_i$ and $\Delta \theta_i$ of these coordinates during the subsequent time-step $\Delta t$ are calculated as
\begin{eqnarray}
\frac{\Delta\mathbf{r}_i}{\Delta t} &=&  u
  \left(\begin{array}{c} \cos\theta_i \\ \sin\theta_i \end{array}\right), \label{updater}\\
\frac{\Delta\theta_i}{\Delta t} &=& {}-\frac{\partial U(\mathbf{r}_1,\dots,\mathbf{r}_N,\theta_1,\dots,\theta_N)}{\partial\theta_i} +\Gamma_i. \label{updatetheta}
\end{eqnarray}
Here, $\Gamma_i$ is a Gaussian stochastic force of zero mean and a correlation of the form $\langle\Gamma_i(t)\Gamma_j(t')\rangle=2D\delta_{ij}\delta(t-t')$. It tends to disturb ordered collective motion. 

The function $U$ contains the alignment rules for the velocity vectors of the particles, 
\begin{eqnarray}\label{U}
\lefteqn{U(\mathbf{r}_1,\dots,\mathbf{r}_N,\theta_1,\dots,\theta_N)} \nonumber\\[.2cm]
&=& {}-\frac{g}{\pi}\sum_{i=1}^{N}\sum_{j=i+1}^{N}\Theta(d_0-\|\mathbf{r}_i-\mathbf{r}_j\|)\cos(\theta_i-\theta_j) \nonumber\\
&& {}-\frac{g_r}{\pi}\sum_{i=1}^{N}\sum_{\substack{j=1,j\neq i}}^{N}\Theta(d_{r}-\|\mathbf{r}_i-\mathbf{r}_j\|) \nonumber\\[-.3cm] 
&& {}\qquad\qquad\qquad\qquad\qquad \times\cos(\theta_i-\measuredangle\frac{\mathbf{r}_i-\mathbf{r}_j}{\|\mathbf{r}_i-\mathbf{r}_j\|}) \nonumber\\
&& {}-\frac{g_w}{\pi}\sum_{i=1}^{N}\Theta(y_i-y_{tw})\cos(\theta_i+\frac{\pi}{2}) \nonumber\\ 
&& {}-\frac{g_w}{\pi}\sum_{i=1}^{N}\Theta(-y_i+y_{bw})\cos(\theta_i-\frac{\pi}{2}), 
\end{eqnarray}
where $\Theta$ denotes the Heaviside step-function. 
First, the term with the coefficient $g>0$ can induce local polar ordering of the velocity vectors of all particles separated by not more than a distance $d_0$. This term constitutes a variant of the original Vicsek model and can enforce global collective motion \cite{peruani2008mean
,menzel2012collective} when $g$ or the mean density is large enough. 

In addition we now include the term with the coefficient $g_r\geq0$: it tends to induce a velocity vector orientation away from the positions of other particles that come closer than $d_{r}<d_0$. (The operator $\measuredangle$ returns the orientation angle of the subsequent unit vector in the two-dimensional plane.) This leads to spatial avoidance and in effect can reduce the local density. A similar approach was suggested in \cite{czirok1996formation}, where, however, an abrupt reorientation rule was introduced instead of a functional form $U$. Other authors used hard \cite{gregoire2003moving} or soft \cite{henkes2011active} potentials to bound the local density. Here, we follow the procedure of a velocity misalignment mechanism, which may be more related to an active decision mechanism of the self-propelling objects. We stress that the exact way of bounding the density is qualitatively irrelevant for the laning and clustering mechanism described in the following, and it can even be obtained without bounding the density.

Furthermore, we now introduce the interaction with two confining walls by the terms with the coefficient $g_w\geq0$. The top wall is located at $y>y_{tw}$, the bottom wall at $y<y_{bw}$, and the distance between the wall surfaces is called $L_y=y_{tw}-y_{bw}$. If a particle penetrates the surface of a wall, its velocity vector tends to align along the normal vector of the wall surface, so that it will leave the wall again. 

We absorb the discrete time-step $\Delta t$ into the parameters on the right-hand side of the equations, $\tilde{g}=g\Delta t$, $\tilde{g}_r=g_r\Delta t$, $\tilde{g}_w=g_w\Delta t$, $\tilde{D}=D(\Delta t)^2$, $\tilde{u}=u\Delta t$, and measure all lengths in units of $\tilde{u}$, $\tilde{\mathbf{r}}_i=\mathbf{r}_i/\tilde{u}$, $\tilde{d}_0=d_0/\tilde{u}$, $\tilde{d}_r=d_r/\tilde{u}$, $\tilde{L}_y=L_y/\tilde{u}$. 
Thus the results for a higher particle speed correspond to those of an affinely contracted system.

\section{Spatial avoidance}\label{repulsive}

The consequences of the term with the coefficient $\tilde{g}_r$ on the collective behaviour are at least twofold. First, the particles are scattered by each other when they come close enough. Due to the many collisions in a many-particle system this effect causes disorder. Therefore the polar alignment of the migration directions is disturbed. Second the system tends to reduce its local density. This becomes important at the transition from disordered to ordered polar collective motion. During a density fluctuation, the density may locally become large enough so that the propulsion directions of the particles within this high-density region align. In this way polar collective motion can be induced. Bounding the local density, the $\tilde{g}_r$-term can inhibit such a scenario. 

\begin{figure}
\centerline{\includegraphics[width=8.5cm]{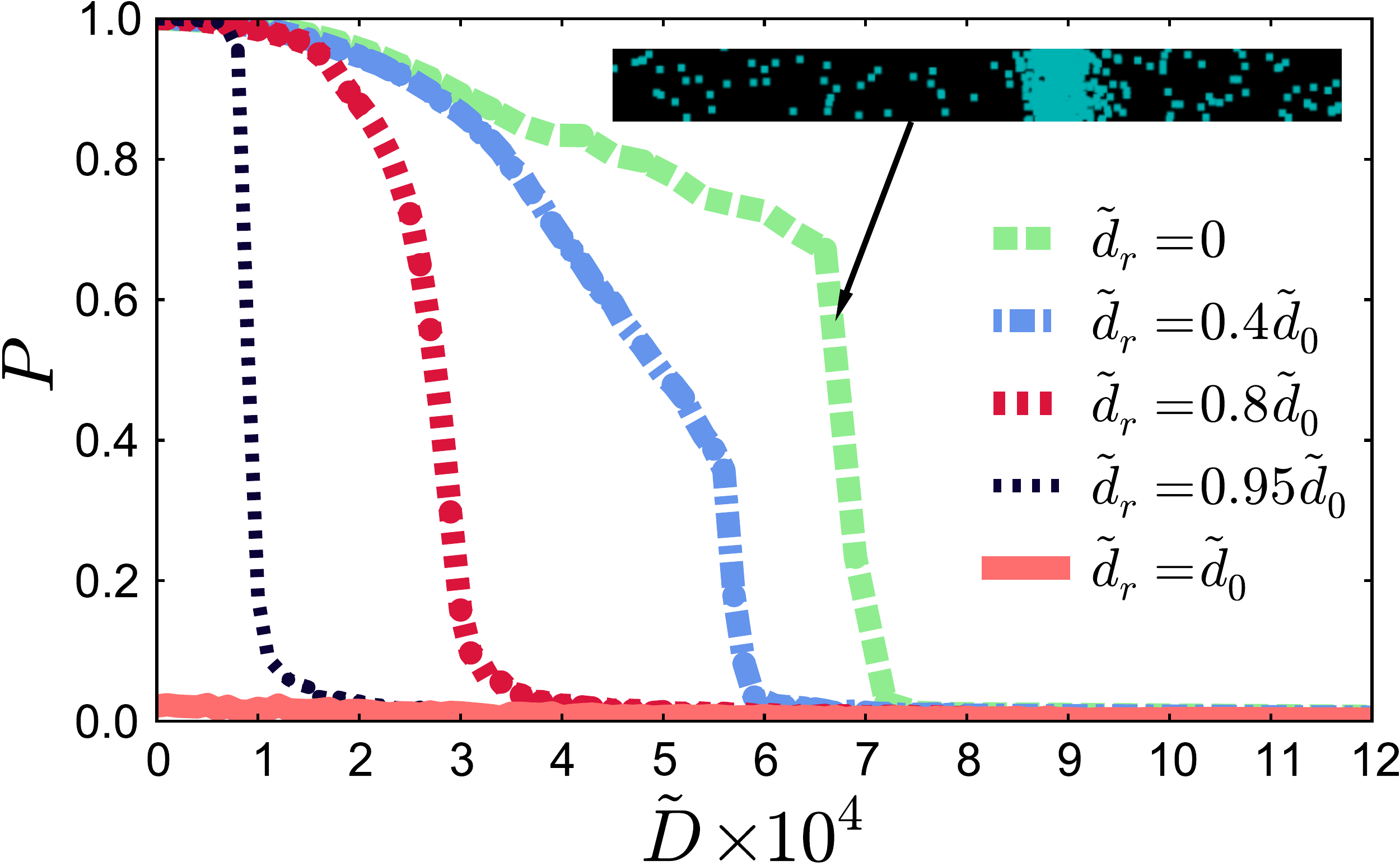}}
\caption{(Colour online). Polar order parameter $P$ of the particle migration directions as a function of the strength of the stochastic force $\tilde{D}$ for different scattering cross sections $\tilde{d}_r$. The measurements were performed in a rectangular calculation box of dimensions $\tilde{L}_x=4000$, $\tilde{L}_y=400$ with periodic boundary conditions and system parameters $\tilde{g}=0.001$, $\tilde{g}_r=0.01$, $\tilde{d}_0=200$, $N=500$. We present time- and sample-averaged values of $P$. The inset shows a high-density band that appears for $\tilde{d}_r=0$ close to the threshold.} 
\label{fig_spatial_avoidance}
\end{figure}

To confirm these effects we investigated a numerical system in a rectangular box with periodic boundary conditions and without the wall terms in (\ref{U}). 
We measured the polar degree of order $P=\|\frac{1}{N}\sum_{i=1}^N\Delta\tilde{\mathbf{r}}_i\|$ of the particle migration directions.  
This order parameter is plotted as a function of the strength of the stochastic force $\tilde{D}$ in figure~\ref{fig_spatial_avoidance} for different scattering cross sections $\tilde{d}_r$. 

Without stochastic noise and scattering, i.e.~at $\tilde{D}=0$ and $\tilde{d}_r=0$, the motion is completely ordered with $P=1$. When the stochastic noise is turned on, $\tilde{D}>0$, the degree of order $P$ decreases and finally becomes zero above a certain threshold value of $\tilde{D}$ \cite{vicsek1995novel,peruani2008mean,peruani2011polar,menzel2012collective}. Increasing the scattering cross section $\tilde{d}_r> 0$ generally reduces the degree of order and the threshold shifts to lower values of $\tilde{D}$. 

Around the threshold, the formation of collectively migrating high-density bands \cite{peruani2011polar,menzel2012collective} is supported by the high aspect ratio of the simulation box (see the inset of figure~\ref{fig_spatial_avoidance} for a snapshot). Increasing the scattering cross section $\tilde{d}_r$ bounds the local particle density. In this way, it hinders the formation of local high-density regions.

\section{Phase behaviour}\label{phases}

In the following we confine the system in a channel between two repulsive walls described by the last two terms in (\ref{U}). 
We systematically increased the channel width $\tilde{L}_y$, and for each channel width the number of contained particles $N$. As a measure for the total particle density we define 
\begin{equation}
\tilde{\rho}_t=N/\tilde{L}_x\tilde{L}_y. 
\end{equation}

\begin{figure}
\centerline{\includegraphics[width=8.5cm]{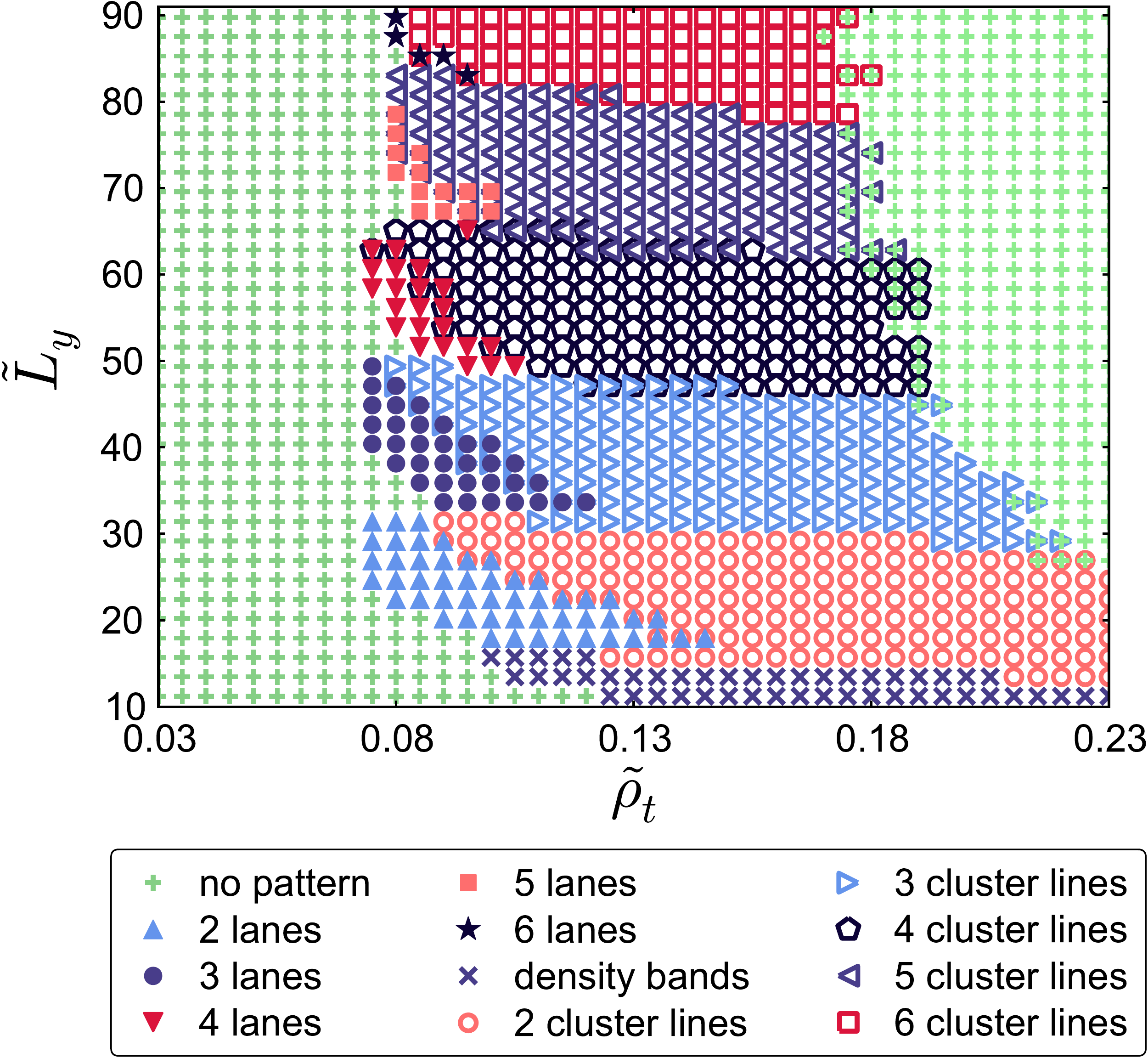}}
\caption{(Colour online). Typical phase behaviour observed for a channel geometry as a function of the total particle density $\tilde{\rho}_t$ and the channel width $\tilde{L}_y$. At low densities $\tilde{\rho}_t$, regular spatial patterns are not observed. When increasing the density, we observed lanes appearing with an elongation parallel to the channel walls. At still higher densities, lines of regularly arranged migrating clusters emerge. If there is only one such line, we call the clusters density bands. The number of lanes or lines of clusters increases with the channel width $\tilde{L}_y$. For very high particle densities $\tilde{\rho}_t$ there were again no regular spatial structures found. The system parameters were set to $\tilde{g}=0.14$, $\tilde{g}_r=0.014$, $\tilde{g}_w=7$, $\tilde{D}=0.004$, $\tilde{d}_0=14$, $\tilde{d}_r=1.4$.} 
\label{fig_phases}
\end{figure}

The resulting phase behaviour in the channel obtained in this way is depicted in figure~\ref{fig_phases} as a function of the total particle density $\tilde{\rho}_t$ and the channel width $\tilde{L}_y$ for the system parameters given in the figure caption. 
The length of the channel $\tilde{L}_x=280$ was always much larger than its width, and periodic boundary conditions were applied at the ends of the channel. 
Increasing the particle number $N$ at a fixed geometry of the system thus corresponds to increasing the particle density $\tilde{\rho}_t$ along the corresponding horizontal line in figure \ref{fig_phases}. 
For each geometry, we started from random initial conditions for the particle positions and velocity orientations and iterated the configuration forward in time (here by $2\times10^5$ time-steps). 

When the system starts to evolve from the random initial conditions, first groups of self-propelled particles form. They collectively migrate and are reflected back and forth between the two confining walls. During this process the crowd is finding the channel elongation as a collective direction of migration, which mainly ceases the collective bouncing process between the channel walls. 

\begin{figure}
\centerline{\includegraphics[width=7.cm]{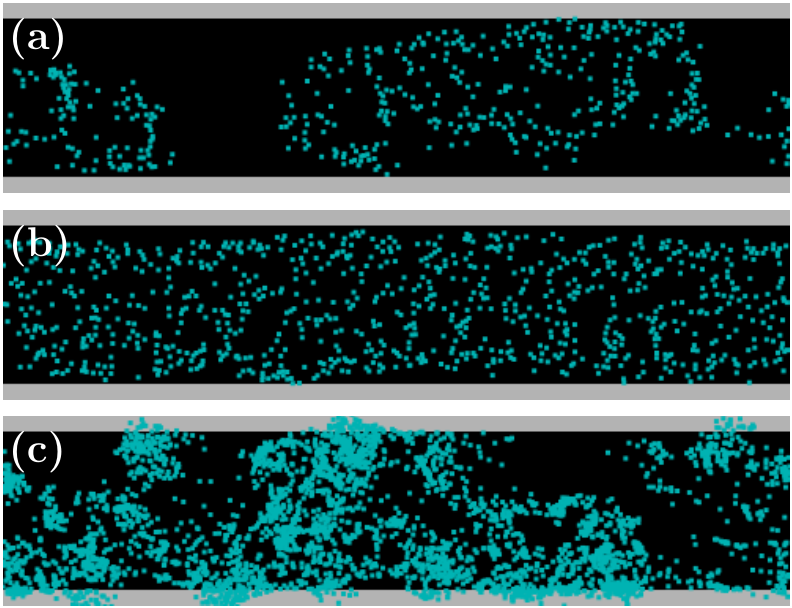}}
\caption{(Colour online). Snapshots of typical low- and high-density scenarios, here at a channel width of $\tilde{L}_y=56$ (all other system parameters are the same as in figure~\ref{fig_phases}). (a) At very low density, here $\tilde{\rho}_t=0.03$, the particle clouds cannot fill the space but find a collective migration direction along the channel. (b) Increasing the density, here to $\tilde{\rho}_t=0.055$, the particle cloud fills the whole space and collectively migrates along the channel. Regular spatial structures do not yet emerge at this density. (c) In contrast to that, at very high densities, here $\tilde{\rho}_t=0.23$, the particle distribution is spatially inhomogeneous, with high-density clouds frequently bouncing between the channel walls, colliding with each other, and as a result sometimes even reversing their sense of migration along the channel.} 
\label{fig_lowhighdensity}
\end{figure}

Figure~\ref{fig_phases} summarises that there are no regular spatial patterns at low particle density $\tilde{\rho}_t$. If the particle density is very low, the cloud of particles cannot fill the space and localised particle clouds appear that migrate along the channel as depicted in figure~\ref{fig_lowhighdensity}~(a). 
Increasing the density makes the cloud more and more fill the whole channel as shown in figure~\ref{fig_lowhighdensity}~(b), with a uniform average migration direction. At some stage of intermediate density $\tilde{\rho}_t$, regular spatially periodic patterns appear. These can be lanes or regularly arranged clusters, both of which are further introduced and analysed in the subsequent sections. 

Finally, at very high densities $\tilde{\rho}_t$, the regular textures disappear again. However, the motion is qualitatively different from the one at low densities. The particle cloud becomes spatially inhomogeneous and does not reach a steady state any more, at least not within the iteration time-spans that we investigated. Instead of that, the motion is rather chaotic. The spatially inhomogeneous particle cloud is continuously bouncing between the channel boundaries. It repeatedly splits into clouds of different migration directions that then collide again due to the boundary conditions of the set-up. In this way, even reversals of the collective migration directions are possible. A snapshot of this state is depicted in figure~\ref{fig_lowhighdensity}~(c).

\section{Unidirectional laning}\label{laning}

\begin{figure}
\centerline{\includegraphics[width=8.5cm]{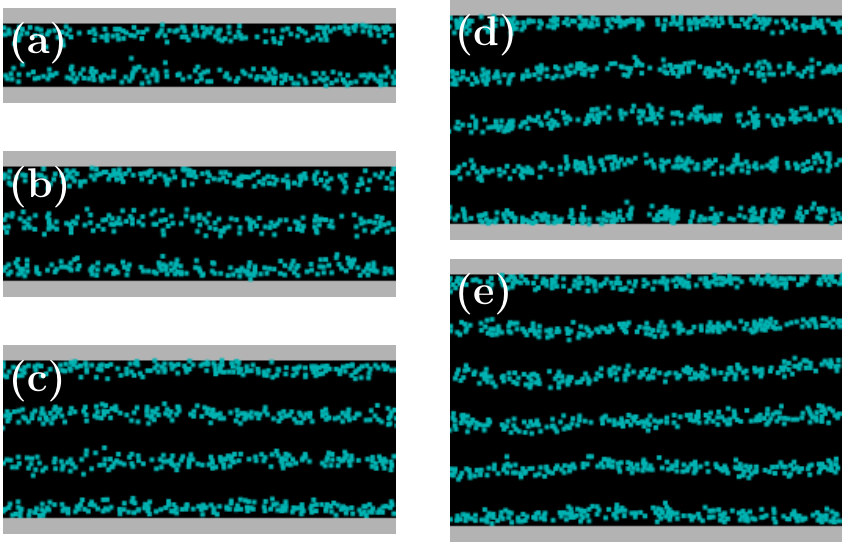}}
\caption{(Colour online). Snapshots of unidirectional laning states emerging for increasing channel width $\tilde{L}_y$. The lanes appear as a steady state from random initial conditions. Within each panel, the particles of all lanes collectively migrate into the same longitudinal direction. We indicate the location of the top and bottom confining walls by grey bars. In all cases the system parameters are the same as in figure~\ref{fig_phases}. The total particle density is fixed at $\tilde{\rho}_t=0.08$ while the channel widths are given by (a) $\tilde{L}_y=22$, (b) $\tilde{L}_y=40$, (c) $\tilde{L}_y=56$, (d) $\tilde{L}_y=74$, (e) $\tilde{L}_y=90$. Only part of the channel is shown in the horizontal direction.} 
\label{fig_laning}
\end{figure}

During the remaining part of this paper, we mainly concentrate on the regular spatial structures appearing at intermediate densities. 
We start with the new laning state that is indicated in the phase diagram figure~\ref{fig_phases} and further depicted in figure~\ref{fig_laning} for different channel widths. 
Iterating the system forward in time from random initial conditions, these lanes appear as a steady final state. They remained stable for as long as we numerically let the system evolve. 

Figure~\ref{fig_laning}~(a) depicts a relatively narrow channel and shows that two lanes emerge that are supported by the confining walls. The space between the lanes is basically empty and is broader than the lane thickness in our example. With increasing channel widths $\tilde{L}_y$, we can find an increasing number of lanes. 
The inner lanes are free-standing in the sense that they are not directly supported by the confining surfaces. We checked that the specific form of the wall boundary conditions is not qualitatively important for the lane formation. Analogous states can also be found from simple reflecting surfaces. 

Former studies reported lanes of self-propelling particles through an environment of immobile particles \cite{henkes2011active}, or neighbouring lanes of opposite migration directions \cite{wensink2012meso, wensink2012emergent, menzel2012soft}. In contrast, all particles in our case have the same speed and migrate collectively on average into the same direction. This is why we introduce the term unidirectional laning. Furthermore, the separation gaps between the lanes are significantly broader in our case than in previous studies \cite{henkes2011active, wensink2012meso, wensink2012emergent, menzel2012soft}. Finally, our particles can also be scattered within the lanes due to the $\tilde{g}_r$-interaction. Altogether, these points indicate a mechanism of lane formation different from the previous studies.

\section{Laning mechanism}\label{mechanism}

\begin{figure}
\centerline{\includegraphics[width=8.5cm]{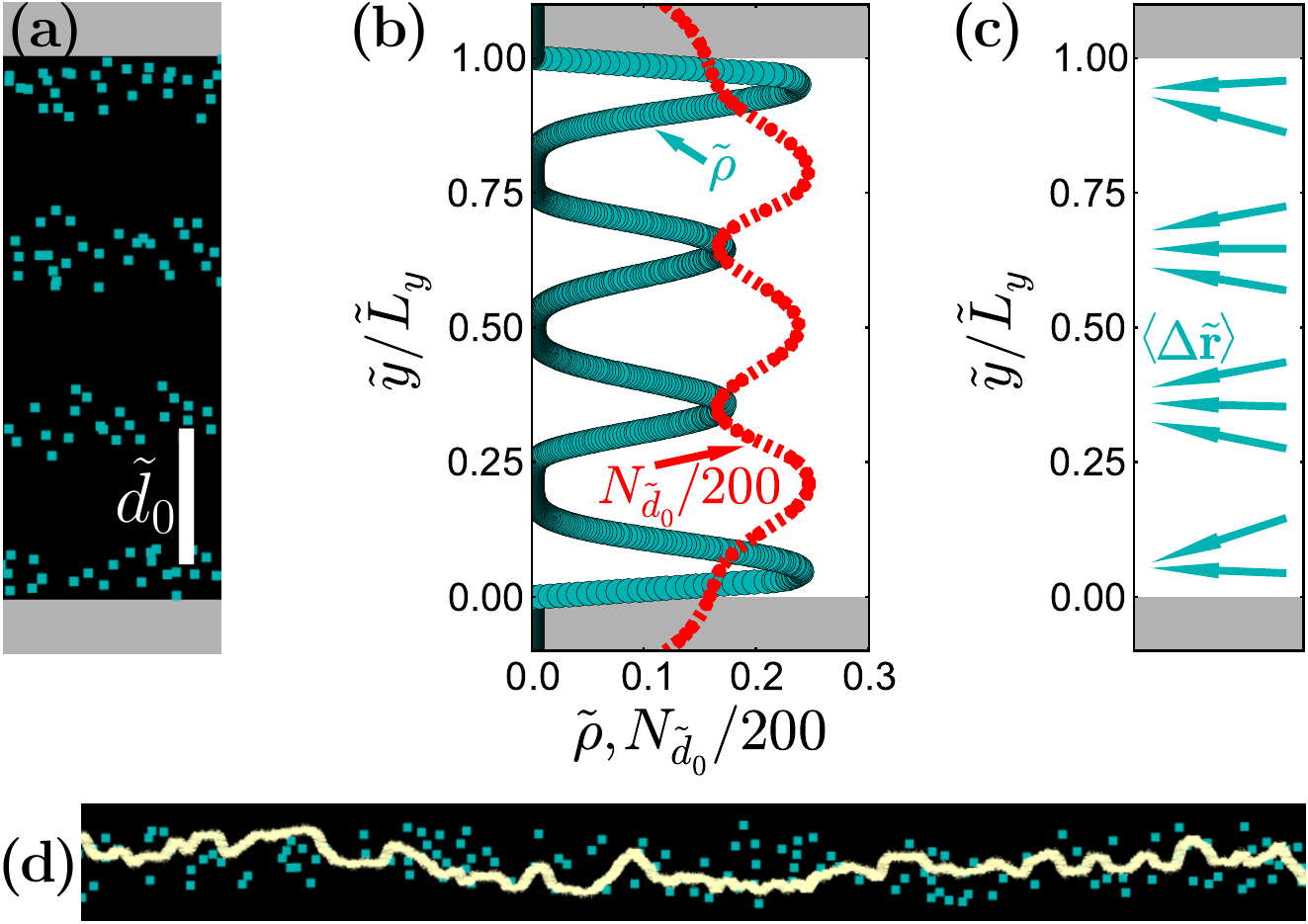}}
\caption{(Colour online). Analysis of the laning state in figure~\ref{fig_laning}~(c). (a)~A magnification of a part of figure~\ref{fig_laning}~(c) shows that the lanes are separated by a distance of order of magnitude $\tilde{d}_0$, as indicated by the white bar. (b)~Across the channel the space- and time-averaged particle density $\tilde{\rho}(\tilde{y}/\tilde{L}_y)$ was evaluated (data points). Here the space-averages are taken in thin horizontal slits along the channel direction. 
We also plot across the channel the local space- and time-averaged number $N_{\tilde{d}_0}(\tilde{y}/\tilde{L}_y)$ of particles within the interaction range $\tilde{d}_0$ (dashed line). (c)~At the positions of the lane centres and lane boundaries the space- and time-averaged directions of the migration steps $\langle\Delta\tilde{\mathbf{r}}\rangle$ are shown. In the $\tilde{y}$-direction the step size is rescaled for better visualization. (d)~An example trajectory of one particle within a free-standing lane demonstrates the inner reflections of the particle off the lane boundaries. The trajectory is contracted in the longitudinal direction to show more reflection events.} 
\label{fig_mechanism}
\end{figure}

To analyse the mechanism behind the formation of the lanes, we first consider the separation distance between the lanes. Figure~\ref{fig_mechanism}~(a) shows a magnification of part of the system in figure~\ref{fig_laning}~(c). 
We infer that the thickness of the gap between the lanes is of the order of the interaction length $\tilde{d}_0=14$. 
If the interaction length $\tilde{d}_0$ is really the determining length scale of the structure, then the multiples of $\tilde{d}_0$ that fit into the channel width $\tilde{L}_y$ should set a lower bound for the number of lanes, taking into account some extra space due to the lane thickness $\tilde{d}_l$. Closer inspection of figure~\ref{fig_phases} demonstrates that this is indeed the case: the areas where we find textures of $n$ lanes ($n-1$ gaps) are systematically bounded below by the channel widths $\tilde{L}_y=(n-1)\tilde{d}_0+n\tilde{d}_l$. Here $n\in\mathbb{N}$ increases from $2$ to $6$ in figure~\ref{fig_phases}, and we obtain $\tilde{d}_l=2$  
in the present case.  

For the system in figure~\ref{fig_laning}~(c) we plot the particle density $\tilde{\rho}$ across the channel in figure~\ref{fig_mechanism}~(b). The density peaks mark the lane positions.  Inner peaks are symmetric with an approximately Gaussian shape whereas the outer peaks become asymmetric due to the wall interactions. To understand the stability of the lanes, we depict in figure~\ref{fig_mechanism}~(c) the predominant migration directions in the lane centres and at their boundaries. It shows that particles reaching the boundaries of a lane mainly reorient back towards the centre of the lane. Or, in other words, for particles that have made a step towards the lane boundary, there is a mechanism that scatters them back into the lane interior. These events of backscattering at the lane boundaries are also evident from the example trajectory of one particle plotted in figure~\ref{fig_mechanism}~(d). It must be this mechanism that stabilises the lane structure. 

Finally, to identify the mechanism, we measured at each position across the channel the number of particles $N_{\tilde{d}_0}$ that are within the interaction range $\tilde{d}_0$. From the plot in figure~\ref{fig_mechanism}~(b) we infer that this number strongly increases at the lane boundaries and is significantly larger between the lanes than within the lanes. This feature follows from the size of the lane separation that is of the order of the interaction range $\tilde{d}_0$. 

Therefore particles within one lane directly interact mainly only with particles in the same lane. As a result their propulsion directions align on average along the lane. 
However, when a particle tries to enter the gap space between the lanes, the number of $\tilde{g}$-interaction partners significantly increases. Now also many particles from the neighbouring lane are within the interaction range $\tilde{d}_0$. The reorientation of the propulsion direction, as it follows from the $g$-term in (\ref{U}), increases with the number of interaction partners. Consequently the angular reorientation of the propulsion direction considerably increases when a particle tries to enter the gap zone. 

In this way, if $\tilde{g}$ is large enough, particles that enter the gap zone overreact. They ``overreorient'' their migration direction back towards the lane. This overreaction stabilises the lanes. For the effect to occur, $\tilde{g}$ must be large enough to induce this overreaction in the gap region, but low enough so that it does not occur significantly within the lanes. Since $\tilde{g}=g\Delta t$, this effect can be tuned by the interaction parameter $g$ or by the size of the time-step $\Delta t$.

\section{Role of the strength of the stochastic force}\label{noiserole}

In section \ref{phases} we analysed the phase behaviour of the system as a function of the channel width $\tilde{L}_y$ and the total particle density $\tilde{\rho}_t$. Our numerical investigations indicate that these are the most immediate parameters to influence the appearance of the ordered structures. However, traditionally the phase behaviour in the Vicsek model is often studied as a function of the magnitude of the orientational noise \cite{vicsek1995novel,chate2008modeling}. Thus we here briefly mention the influence of the corresponding strength of the stochastic force $\tilde{D}$ on the observed textures. 

\begin{figure}
\centerline{\includegraphics[width=8.5cm]{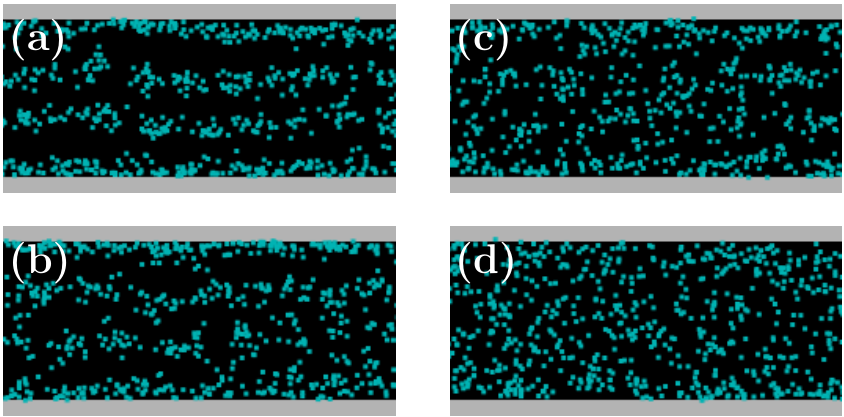}}
\caption{
(Colour online). Snapshots of the unidirectional laning state from figures~\ref{fig_laning} (c) and \ref{fig_mechanism} for increasing strength of the stochastic force $\tilde{D}$. Instead of the previous magnitude $\tilde{D}=0.004$, we here depict steady states at (a) $\tilde{D}=0.012$, (b) $\tilde{D}=0.016$, (c) $\tilde{D}=0.02$, and (d) $\tilde{D}=0.04$. The ordered structures are dissolved at high noise levels. Only a fraction of the channel is shown in the horizontal direction.} 
\label{figDqual}
\end{figure}

As one might expect, an increasing strength of the stochastic force $\tilde{D}$ makes the structures become more fuzzy. In particular, the clear separation of the system into the longitudinal lanes and the empty gap space between these lanes becomes less well-defined. We illustrate this in figure~\ref{figDqual}. 
The system there is identical to the one shown in figure~\ref{fig_laning} (c) and further analysed in figure~\ref{fig_mechanism}, with the only difference that the noise strength $\tilde{D}$ is increased. As a result, the lanes become less pronounced and finally are dissolved. 

\begin{figure}
\centerline{\includegraphics[width=8.5cm]{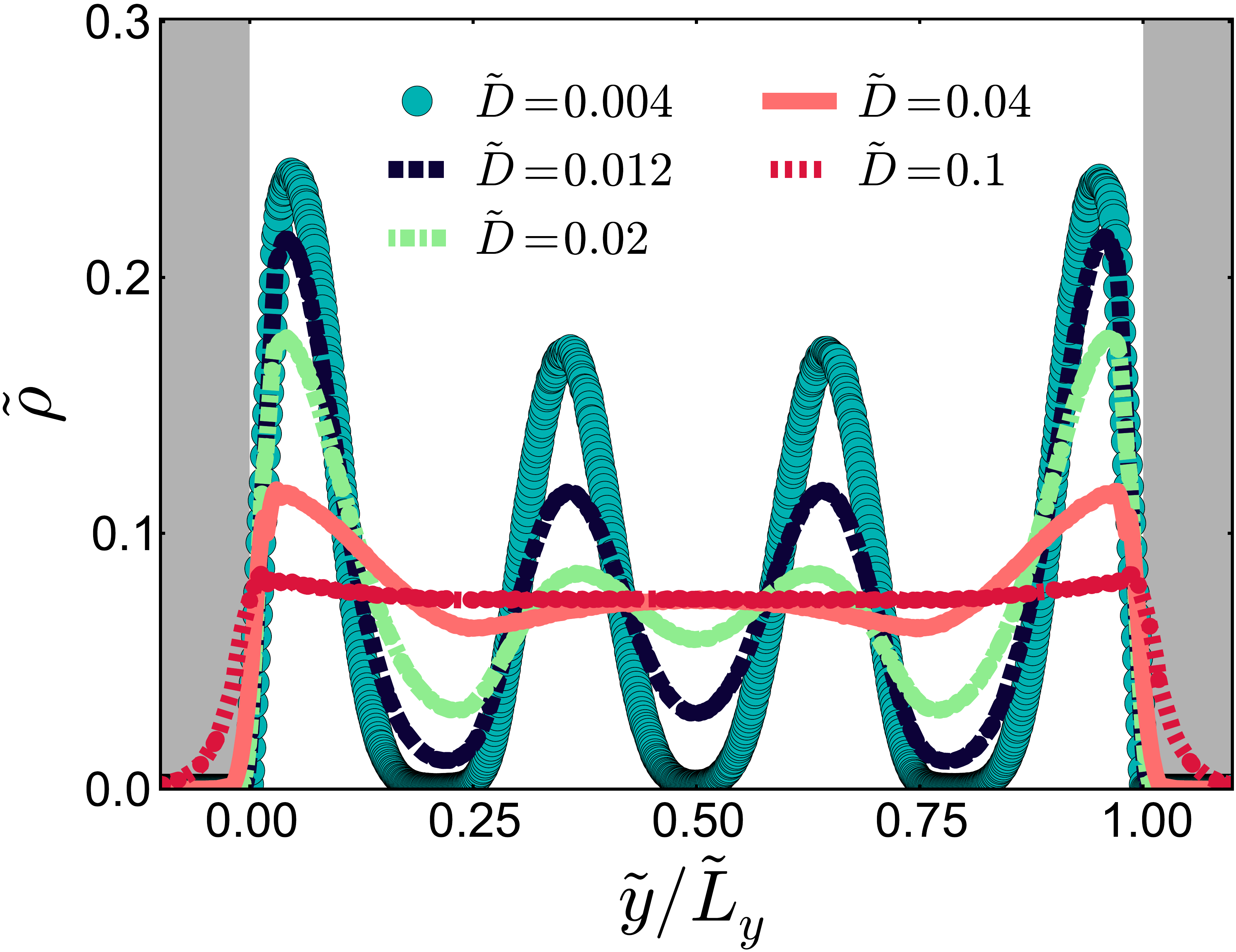}}
\caption{
(Colour online). Space- and time-averaged particle density $\tilde{\rho}(\tilde{y}/\tilde{L}_y)$ across the channel for different strengths $\tilde{D}$ of the stochastic force. The space-averages were taken in thin horizontal slits along the channel direction. An increasing strength of the stochastic force reduces the magnitude of the density modulation.} 
\label{fig_D_quant}
\end{figure}

A more quantitative picture of this scenario is given in figure~\ref{fig_D_quant}. 
There we plot the space- and time-averaged particle density $\tilde{\rho}(\tilde{y}/\tilde{L}_y)$ across the channel as in figure~\ref{fig_mechanism} (b), but here for different increasing values of $\tilde{D}$. For visualization, the plot is rotated by ninety degrees when compared to figure~\ref{fig_mechanism} (b). With increasing magnitude of $\tilde{D}$, the peaks shrink in height. Furthermore, the gap regions between the peaks are more and more filled with particles. In all cases, the outer peaks are more pronounced that the inner ones. This is a consequence of the supporting boundary conditions. At moderate noise levels, the number of lanes and their thicknesses remain basically constant. At high noise levels, the structure is finally dissolved. 

In addition to that, we checked how the phase boundaries in figure~\ref{fig_phases} are affected. We concentrated on moderate noise levels $\tilde{D}$ that do not dissolve the ordered structures. For the cases analysed and within the resolution of figure~\ref{fig_phases}, we did not observe a noise-induced shift of the boundaries separating the different states. 

\section{Migrating clusters}\label{clusters}

\begin{figure}
\centerline{\includegraphics[width=8.5cm]{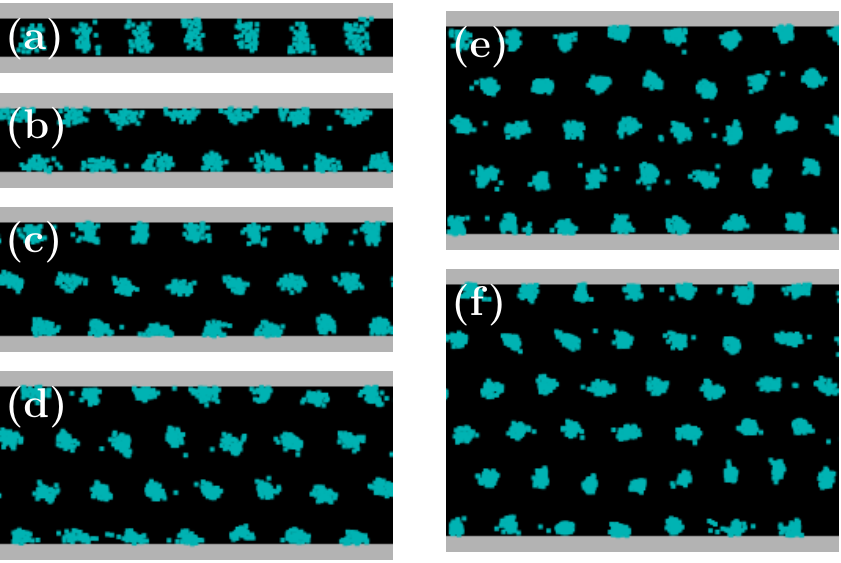}}
\caption{(Colour online). Snapshots of steady states of collectively migrating density bands (a) and clusters (b)--(f). All clusters migrate on average into the same direction, thus maintaining the approximately hexagonal order in panels (c)--(f). The total particle density is fixed at $\tilde{\rho}_t=0.13$, whereas the channel widths vary from (a) $\tilde{L}_y=14$, (b) $\tilde{L}_y=22$, (c) $\tilde{L}_y=40$, (d) $\tilde{L}_y=56$, (e) $\tilde{L}_y=74$, to (f) $\tilde{L}_y=90$.} Since all other parameters are the same as in figure~\ref{fig_phases}, panels (b)--(f) characterise the systems of figure~\ref{fig_laning}~(a)--(e) at a higher total particle density, respectively. Only a fraction of the channel is shown in the horizontal direction. 
\label{fig_cluster}
\end{figure}

Generally we do not observe the lanes in figure~\ref{fig_phases} any more at higher particle densities $\tilde{\rho}_t$. Instead, the system becomes spatially modulated also in the longitudinal direction. We find neighbouring lines of migrating clusters emerging. Examples are depicted in figure~\ref{fig_cluster}. 

Within a cluster, the velocity vectors of the particles collectively align so that the cluster propels as one unit object. The resulting crowd of clusters, and therefore all particles together, are collectively migrating on average into the same direction along the channel. 

We start with the case of thin channels as depicted in figure~\ref{fig_cluster}~(a). There we can observe a line of approximately equally spaced traveling clusters that are extended between the channel walls. 
We termed these objects density bands in figure~\ref{fig_phases}. 
Apart from the ingredients that lead to the traveling high-density bands observed in the Vicsek model \cite{chate2008modeling,ihle2013invasion}, here the mechanism that we identified as inducing laning is at work as well and promotes the modulation in the longitudinal direction. 
Furthermore, the formation of the clusters in our case is strongly supported by the confining walls. 

In broader channels, we can observe several neighbouring lines of migrating clusters as shown in figure~\ref{fig_cluster}~(b)--(f). 
The number of horizontal cluster lines corresponds to the number of lanes observed at lower density $\tilde{\rho}_t$. Likewise we find free-standing lines of migrating clusters that are not directly supported by the walls. Furthermore, the clusters themselves are arranged on an approximately hexagonal lattice. 
(Due to the hexagonal arrangement, neighbouring cluster lines are shifted with respect to each other in the channel direction. Therefore a characteristic distance $\tilde{d}_0$ between the clusters can be adopted at slightly thinner channels than for the lane structures. This might be the reason why in figure~\ref{fig_phases} clusters appear at slightly lower channel widths $\tilde{L}_y$ than the corresponding lane textures.)

On the one hand, active but resting cluster crystals were previously found for a system of deformable self-propelled particles \cite{menzel2012soft} and in a broader sense also in systems of circle swimmers \cite{riedel2005self,kaiser2013vortex}. On the other hand, collectively traveling crystalline textures in non-equilibrium systems were reported before \cite{okuzono2001self,sugiura2002time,okuzono2003traveling,gregoire2003moving,menzel2013traveling,ferrante2013elasticity}. Here we observe the combination of these two situations, i.e.\ traveling active cluster crystals -- albeit supported by the confining boundaries. 

\begin{figure}
\centerline{\includegraphics[width=8.5cm]{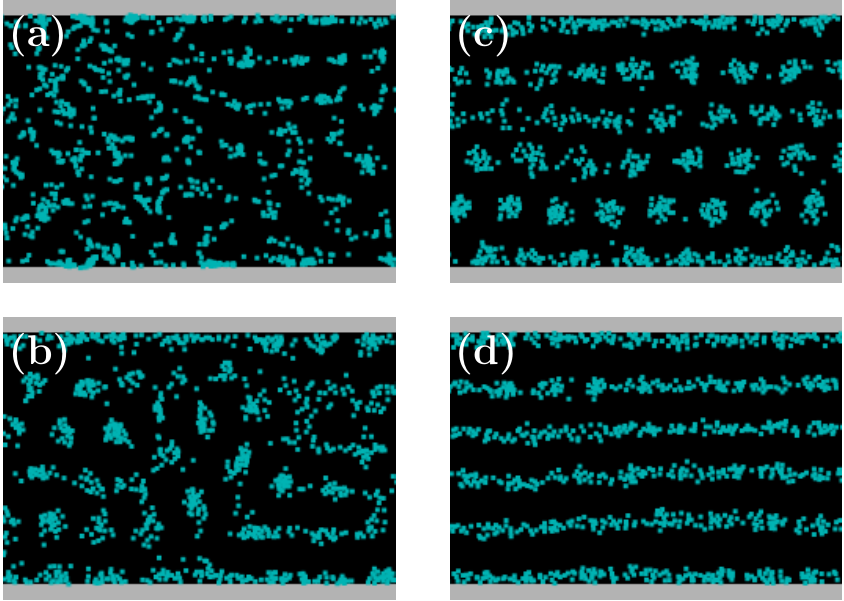}}
\caption{(Colour online). Transition from migrating clusters to lanes during the formation of the unidirectional laning state as depicted in figure~\ref{fig_laning}~(e). The snapshots were taken while the texture was evolving from random initial conditions. We depict its state after (a) 300, (b) 3000, (c) 30000, and (d) 300000 time-steps. Only part of the channel is shown in the horizontal direction. }
\label{fig_evolution}
\end{figure}

Furthermore we note that the clusters can also emerge as an intermediate state during the formation of the unidirectional laning textures. We demonstrate such a scenario in figure~\ref{fig_evolution}. 
There, migrating clusters form from the random initial conditions over time. They order in a hexagonal way. Only at later evolution times these clusters smoothen out until finally the laning texture appears. We checked, however, that the migrating cluster crystals in figure~\ref{fig_cluster} remain stable even after very long iteration times.

\section{Role of the misalignment rule}\label{role}

In the previous sections, we bounded the local density by applying the rule of velocity reorientation given by the term with the coefficient $g_r$ in (\ref{U}). We stress here that this effectively repulsive mechanism is not a necessary ingredient to obtain the regular textures that we observed. Unidirectional laning and migrating clusters are also found without this interaction, i.e.\ for $g_r=0$. 

We further checked this point by applying a different misalignment rule for the velocity vectors. In this other case, if particles come closer than a distance $d_{\perp}<d_0$, they tend to align their velocity vectors perpendicular to each other. This can be achieved via the contribution
\begin{equation}
{}-\frac{g_{\perp}}{\pi}\sum_{i=1}^{N}\sum_{j=i+1}^{N}\Theta(d_{\perp}-\|\mathbf{r}_i-\mathbf{r}_j\|)\sin^2(\theta_i-\theta_j) 
\end{equation}
instead of the $g_r$-term in (\ref{U}). 
The interaction effectively bounds the particle density. 

\begin{figure}
\centerline{\includegraphics[width=6.cm]{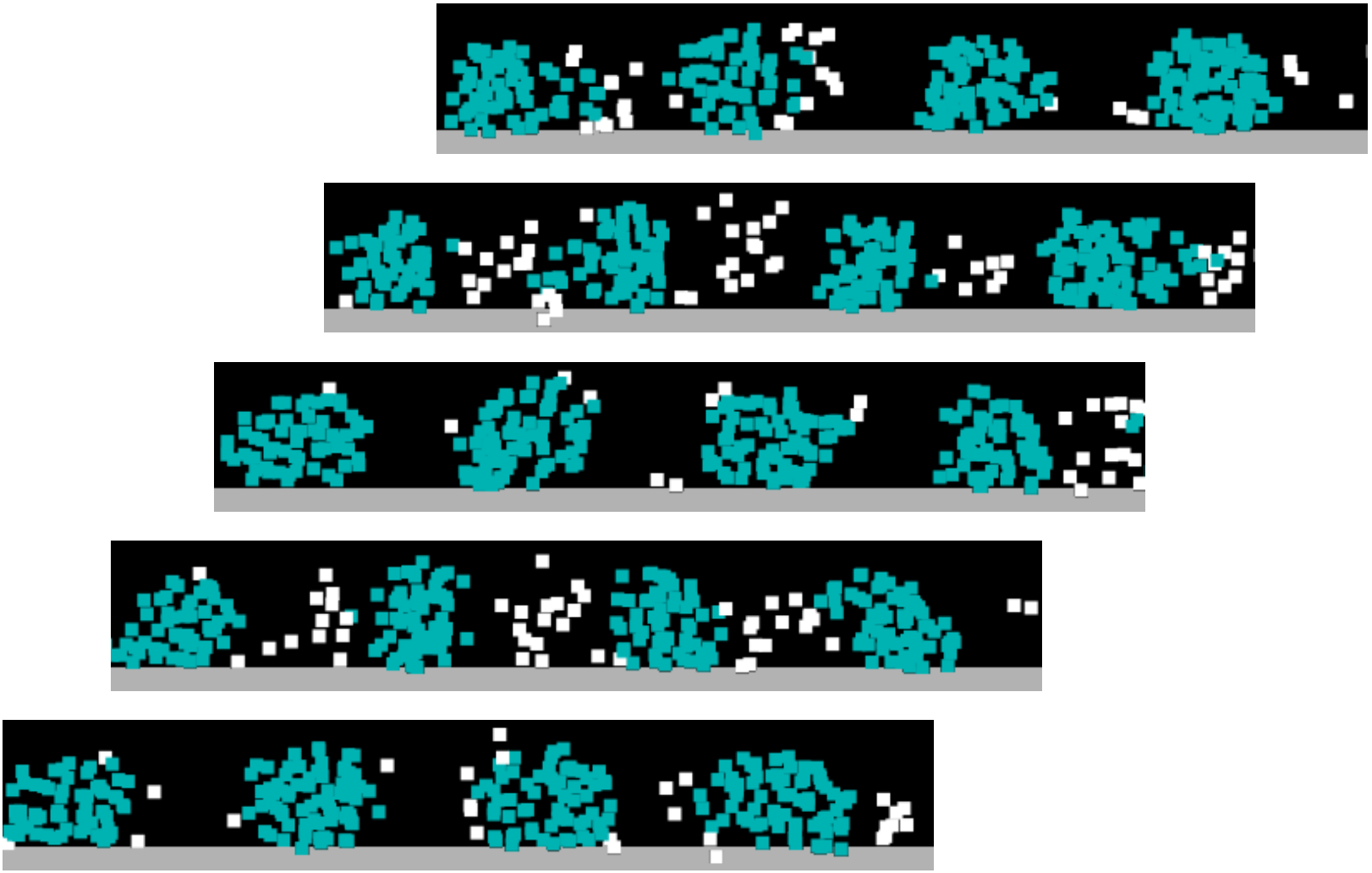}}
\caption{(Colour online). Illustration of the swapping process. From top to bottom a time series of four clusters is shown that migrate to the left. Particles that are not organised within a compact cluster and therefore have a reduced velocity component along the channel are shown in white. Such particle clouds are continuously left behind by one cluster and picked up by the next cluster. The panels correspond to zoomed snapshots onto the lower channel boundary of a system analogous to the one in figure~\ref{fig_cluster}~(b), but with a perpendicular alignment rule for the velocity vectors replacing the repulsive one of the $\tilde{g}_r$-interaction. System parameters: $\tilde{L}_y=38$, $\tilde{\rho}_t=0.15$, $\tilde{g}=0.1$, $\tilde{g}_{\perp}=1$, $\tilde{g}_w=5$, $\tilde{D}=0.002$, $\tilde{d}_0=20$, $\tilde{d}_{\perp}=2$.} 
\label{fig_swapping}
\end{figure}

Again we observe the lanes and migrating clusters in certain parameter regimes.
In addition, we found for this case that the collectively migrating clusters feature a dynamic effect that we term swapping and that we depict in figure~\ref{fig_swapping}. 

The compact clusters shown in this figure are collectively migrating to the left. Repeatedly the directions of self-propulsion of some particles at the rear of a cluster become disordered with respect to the collective migration direction. As a consequence these particles have a larger velocity component perpendicular to the channel and therefore migrate more slowly along the channel. Thus they cannot follow the cluster any more. 

A cloud of such particles is therefore left behind. This excess material is then picked up by the next cluster and the process continues. 
In this way the particles are passed through the line of successive clusters. Within the comoving frame of the clusters, it looks as if some excess material swaps continuously from one cluster to the next. 

Altogether we can say that the regular textures are obtained \textit{although} we applied the repulsive interaction. Generally, the structures dissolve with increasing strength $\tilde{D}$ of the orientational noise. A similar effect follows from the scattering of the particles from each other due to the misalignment rules for the velocity vectors.

\section{Discussion and conclusions}\label{conclusion}

In this study we investigated a variant of the Vicsek model that uses a functional form to define the velocity alignment rules \cite{peruani2008mean,menzel2012collective}. An additional particle scattering reduces the agglomeration of particles in high-density regions. We found that in a confining channel an overreaction in velocity alignment can induce unidirectional laning. This means that all particles in all lanes on average propel into the same direction. The lanes are separated by a finite gap. 
For other parameter ranges we observed lines of clusters migrating into the same direction. These clusters arrange in an approximately hexagonal spatial order. 

Previous studies have reported an accumulation of self-propelled particles at surfaces \cite{wensink2008aggregation,costanzo2012transport,elgeti2013wall}. However, in these studies the mechanism of accumulation is different. The particles are driven by a constant propulsion force. When a particle is driven towards a repelling surface, it is slowed down. Other particles may block this first particle and hinder its reorientation. In this way the possibility to leave the surface can be significantly decreased. In contrast, in our case the particles always propel with constant velocity magnitude. When they hit a surface, they are even supported to leave it again: through the wall interaction, their propulsion directions are reoriented away from the surface. A certain number of lanes in our case emerges when the range of the alignment interaction matches a certain ratio of the channel width. The mechanism behind the formation of the lanes and the depletion zones between them is an overreaction in velocity reorientation when a particle tries to enter the depletion zone. As a consequence the particles are hindered from crossing the depletion zones. In this way we can also observe free-standing lanes that are not directly supported by an adjacent surface. 

Finally, we conclude by two remarks. First, to obtain the presented results, we made all self-propelled particles simultaneously update their migration directions and positions. However, we checked qualitatively the reproducibility also for a non-simultaneous update procedure. And second, as we explained, a major part of our results rely on the effect of a discrete time-step. This is not as artificial as it might appear at first glance, for example when we think of real self-propelling objects in nature. Birds and insects fly by discrete flaps of their wings, fish swim by discrete strokes of their fins, humans and land animals move by discrete steps or jumps. Before each step or jump, the direction of motion has to be decided. It is difficult to correct this direction during the discrete migration step. Taking these points together, we are optimistic that our results can be observed also in real systems of self-propelling objects.

\ack{
The author thanks Hartmut L\"owen, Takao Ohta, Harald Pleiner, Hirofumi Wada, as well as Mitsusuke Tarama for stimulating discussions, and the Deutsche Forschungsgemeinschaft for partial support of this work through the German-Japanese project ``Nichtgleichgewichtsph\"anomene in Weicher Materie/Soft Matter'' LO 418/15. 
}

\end{document}